# Molecular Beam Epitaxy of 2D-layered Gallium Selenide on GaN substrates

*Choong Hee Lee[1,*], Sriram Krishnamoorthy[1,*], Dante J. O'Hara[2], Jared M. Johnson[3], John Jamison[3], Roberto C. Myers[3], Roland K. Kawakami[2,4], Jinwoo Hwang[3], Siddharth Rajan[1]*

* *Choong Hee Lee and Sriram Krishnamoorthy have contributed equally to this work.*

[1] Department of Electrical and Computer Engineering, The Ohio State University, Columbus, Ohio 43210, USA

[2] Program of Materials Science and Engineering, University of California, Riverside, California 92521, USA

[3] Department of Materials Science and Engineering, The Ohio State University, Columbus, Ohio 43210, USA

[4] Department of Physics, The Ohio State University, Columbus, OH 43210, USA

KEYWORDS

GaSe, metal chalcogenides, molecular beam epitaxy, van der Waals epitaxy

ABSTRACT

Large area epitaxy of two-dimensional (2D) layered materials with high material quality is a crucial step in realizing novel device applications based on 2D materials. In this work, we report high-quality, crystalline, large-area gallium selenide (GaSe) films grown on bulk substrates such as *c*-plane sapphire and gallium nitride (GaN) using a valved cracker source for Se. (002)-oriented GaSe with random in-plane orientation of domains was grown on sapphire and GaN substrates at a substrate temperature of 350-450 $^o$C with complete surface coverage and smooth surface morphology. Higher growth temperature (575 $^o$C) resulted in the formation of single-crystalline ε-GaSe triangular domains with six-fold symmetry confirmed by *in-situ* reflection high electron energy diffraction (RHEED) and off-axis x-ray diffraction (XRD). A two-step growth method involving high temperature nucleation of single crystalline domains and low temperature growth to enhance coalescence was adopted to obtain continuous (002)-oriented GaSe with an epitaxial relationship with the substrate. While six-fold symmetry was maintained in the two step growth, β-GaSe phase was observed in addition to the dominant ε-GaSe in cross-sectional scanning transmission electron microscopy images. This work demonstrates the potential of growing high quality 2D-layered materials using molecular beam epitaxy and can be extended to the growth of other transition metal chalcogenides.

INTRODUCTION

Two-dimensional (2D) metal chalcogenides are of great scientific interest for electronic as well as optical devices due to their unique structural, electrical and mechanical properties such as wide range of bandgaps[1-2], valley-polarized carriers[3-4], strong spin–orbit coupling[5], and superconductivity[6]. Recently, artificial stacking of these layered materials is being heavily explored to create heterostructures for novel applications. Most of these studies have been carried out by transferring layered flakes or films from minerals[7-9] or synthesized materials obtained using chemical vapor transport (CVT)[10-11] or chemical vapor deposition (CVD) methods [12-13]. In contrast to such stacking methods, epitaxial techniques such as metal organic chemical vapor deposition and molecular beam epitaxy (MBE) provide a more practical approach to achieving large area epitaxial materials with precise control of layer thickness and doping. In addition, the absence of out-of-plane dangling bonds in layered materials can enable van der Waals epitaxy (vdWE) on highly lattice-mismatched substrate without lattice matching constraints.[14,15] For instance, growth of gallium selenide on mica substrates with a significant lattice mismatch of 35% has been reported.[14]

Van der Waals epitaxy was first introduced by Koma and co-workers[15] and has been proven to be a powerful route to realize heteroepitaxy of 2D materials. More recently, renewed interest in 2D materials has led to exploration of MBE growth of several materials including $GaSe$[16-17], $MoSe_2$[18-22], $WSe_2$[23] and $HfSe_2$[24]. In this work, we report our work on growth of gallium selenide (GaSe), which has a layered crystal structure consisting of repeating units of covalently bonded Se-Ga-Ga-Se held together by weak van der Waals force.

Layered GaSe, however, occur in several polytypes displaying different stacking sequences, leading to $\varepsilon$-, $\beta$-, $\gamma$-, and $\delta$- phases of the material.[25] Most common polytypes, $\varepsilon$ (consist of two layers per unit cell and has the space group, $D^1_{3h}$) and *β* (consist of two layers and has the space group, $D^4_{6h}$), have a 2H staking sequence.[26] Bulk *ε*-GaSe is a 2 eV direct bandgap semiconductor and has been explored for applications in nonlinear optics, photovoltaics and photodetectors.[27-28]

Single crystal MBE growth of GaSe on GaAs(111)B substrates has been reported by Keiji *et al*.[29] It has also been shown that GaSe and $Ga_2Se_3$ can be grown on GaAs (001) substrates depending on the surface reconstruction.[30] Vinh *et al.* demonstrated the growth of single crystal GaSe film on Si(111) substrate with 7×7 surface reconstruction.[31] In addition, recent studies report that the growth of GaSe on sapphire substrates produces crystalline films with random in-plane orientation of the domains.[16]

However, there have not been reports on GaSe growth on wide bandgap semiconductors such as gallium nitride (GaN). Epitaxially grown high quality 2D materials on GaN can enable vertical 2D/3D heterostructures[32-33] that can enable vertical tunneling devices[33], heterojunction bipolar transistors (HBT), and hot electron transistors. We demonstrate the growth of highly crystalline centimeter-scale few layer GaSe films on bulk 3D materials such as sapphire and GaN. First, we have investigated the growth of continuous GaSe film on sapphire substrates at various growth conditions and utilized the optimized condition to grow GaSe on a GaN substrate. We report a two-step growth method to grow crystalline GaSe on GaN substrates by employing a high temperature nucleation step for growth of single crystal domains followed by the second step at lower growth temperature to achieve coalescence of the film.

RESULTS AND DISCUSSION

Growth of GaSe was explored on *c*-plane sapphire substrates by varying the substrate temperature and the Ga:Se flux ratio. *C*-plane sapphire was chosen due to the hexagonal symmetry of the basal plane, which is similar to that of GaSe, and the high chemical and thermal stability of sapphire. The substrate temperature was varied from 350°C to 500° C, while changing the Ga:Se ratio from 1:50 to 1:200, and holding the Se flux at $1\times10^{-6}$ Torr. Growth was performed for one hour. The Se shutter was opened for two minutes and the streaky RHEED pattern of sapphire substrates (Fig. 1(a)) remained before the opening of the Ga shutter indicating that the sticking of Se adatoms is very poor in the absence of Ga flux. Upon opening the Ga shutter, the RHEED pattern corresponding to *m*- ($10\bar{1}0$) and *a*- ($11\bar{2}0$) planes of GaSe was observed, and the RHEED pattern did not change along the different azimuths (i.e. in-plane rotation of the substrate). The coexistence of the RHEED patterns corresponding to the *m*- and *a*-planes of GaSe was also reported earlier.[16] This indicates that GaSe nucleated with random in-plane orientation. However, no polycrystalline rings were observed in the RHEED. The inverse of the ratio of spacing between *m*-plane and *a*-plane streaks in the RHEED image (Figure 1(b)) was measured to be 1.72, which is very close to the theoretical value of $\sqrt{3}$. XRD spectra of the samples grown in the range of conditions mentioned above, with the exception of the extremely Se-rich condition ($T_{sub}$ = 400 °C, Ga:Se = 1:200) showed diffraction peaks corresponding to the (002) family of planes in layered-GaSe. However, with extremely Se-rich condition, the $Ga_2Se_3$ phase was observed in XRD and a spotty RHEED pattern was observed (Fig. S1 in the supplementary information (SI)). The growth window for GaSe in order to maintain a streaky RHEED pattern was found to be very narrow at a given substrate temperature. The RHEED pattern remained streaky and the intensity remained constant only at a certain Ga flux at a given substrate temperature; Higher

Ga flux resulted in complete RHEED dimming and lower Ga flux resulted in a spotty RHEED pattern.

Surface morphology of GaSe films as a function of growth conditions is shown in figure 2(a). At 350 $^o$C, a Ga:Se ratio of 1:50 resulted in a smooth surface morphology, while a reduced Ga flux (Ga:Se = 1:100) was required at a growth temperature of 400 $^o$C. With an increase in substrate temperature from 350 $^o$C to 450 $^o$C, smooth surface morphology could be maintained only with a reduction of Ga flux. This can be explained as follows. With an increase in substrate temperature, the sticking coefficient of Se is expected to reduce and hence the Ga flux that is required to maintain stoichiometry at the surface is lower at higher substrate temperatures. This is also expected to result in a reduction in the growth rate of GaSe with an increase in the substrate temperature, assuming unity sticking coefficient for Ga adatoms at the growth temperature used. This observation is in agreement with the RHEED patterns observed during the growth. At the optimized conditions, where the adsorbed Ga and Se adatoms are close to stoichiometry, the RHEED pattern remained streaky throughout the growth. However, when the Ga flux is higher than the stoichiometry (Ga:Se =1:50, $T_{sub}$ = 400 $^o$C, 450 $^o$C) the RHEED showed an amorphous pattern indicating the presence of excess Ga on the surface during the growth. With Se-rich conditions, a spotty (i.e. rough) RHEED pattern as observed. At higher substrate temperatures (>500 $^o$C), the Se sticking coefficient is very low and no growth was observed. At the optimized conditions with streaky RHEED pattern and bright RHEED intensity, a very smooth surface morphology (rms roughness = 0.76 nm) with atomic steps was observed (Fig. 2(b)), indicating step flow growth regime. The step height measured from AFM (0.8 nm) matches closely with the thickness of monolayer GaSe.

Using the optimized growth conditions obtained from growth studies on sapphire ($T_{sub}$ = 400 $^o$C, Ga:Se 1:100), we next explored the growth of GaSe on GaN templates (2 μm GaN/sapphire). The films were grown for one hour and the growth rate was found to be 0.75 nm/min with a total film thickness of 45 nm. While the lattice mismatch between GaN and GaSe (18%) is high, GaN provides a direct route for device design using 2D/GaN heterostructure based devices.

X-ray diffraction of the films (Fig. 3(b)) showed diffraction peaks corresponding to (002), (004), (006) and (008) planes of GaSe. Complete surface coverage with a very smooth surface morphology (rms roughness = 0.85 nm) with spiral hillocks and atomic steps, indicative of step flow growth regime, was obtained. However, the sample showed in-plane disorder (Fig. 3(a)) showing both *m*-plane and *a*-plane spacing, and the RHEED pattern was insensitive to substrate rotation.

Control of in-plane orientation of the crystal domains during nucleation is very critical to obtaining single crystalline GaSe films. GaSe growth temperature was hence increased from 400 $^o$C to 575 $^o$C to control the in-plane orientation. Higher growth temperature necessitates higher Se flux due to reduction in sticking coefficient of Se with increase in the substrate temperature. The Se beam flux was increased to $1 \times 10^{-5}$ Torr.

Figure 4(a) shows XRD spectra of GaSe films grown at different growth conditions on GaN substrates. The sample grown at 500 $^o$C with at a ratio of 1:400 showed (111) oriented $Ga_2Se_3$ phase due to excess Se. With an increase in Ga:Se ratio from 1:400 to 1:100 diffraction peaks corresponding to both GaSe(002) and $Ga_2Se_3$(111) planes were measured. With further increase in Ga:Se ratio to 1:100, only GaSe(002) was detected at higher growth temperature of 575 $^o$C.

Figure 4(b)-(e) show the RHEED patterns of GaN substrate and GaSe film grown at 575 $^o$C with Ga:Se = 1:100 along the [11$\bar{2}$0] and [10$\bar{1}$0] directions of the GaN substrate. The RHEED patterns corresponding to *m*- and *a*-planes of GaSe (Fig. 4 (d) and (e), respectively) were observed along the same azimuth as GaN. The basal planes of GaSe was found to be perfectly aligned with the GaN substrate ([11$\bar{2}$0]GaSe//[11$\bar{2}$0]GaN and [10$\bar{1}$0]GaSe//[10$\bar{1}$0]GaN) and six-fold symmetry of GaSe was clearly observed. Unlike the film grown at 400 $^o$C with in-plane disorder, GaSe streaks corresponding to *m*- and *a*-planes of GaSe appeared only at every 60$^o$ azimuthal rotation spacing. The inverse of the RHEED spacing ratio between GaN and GaSe was found to be 1.170, which is very close to the ratio (1.173) of bulk lattice constants of GaSe (3.74 nm) and GaN (3.189 nm). This clearly suggests that the epilayer is fully relaxed and the growth proceeds by van der Waals epitaxy.

While the higher temperature growths led to single phase films, surface coverage was found to be incomplete. A step height corresponding to 1 ML of GaSe (0.8 nm) was measured at the edge of a triangular domain that grew on top of another triangular domain. Large area (10 μm × 10 μm) AFM scan (Fig. S2 in SI) and STEM measurements (Fig. S3 in SI) confirmed the observation of incomplete surface coverage from AFM scans. More details regarding the microstructure of the film is discussed later in the manuscript.

While high temperature growth of GaSe at 575$^o$C resulted in (002)-oriented single crystal domains, the layers did not coalesce to form a continuous layer. Growth at 400 $^o$C with a Ga:Se ratio of 1:100 resulted in smooth (002)-oriented GaSe layers with in-plane disorder. To obtain single crystalline GaSe with complete surface coverage, we designed a two-step growth method illustrated in Fig. 5(a). After forming the nucleation layer at 575 $^o$C with 1×10$^{-5}$ Torr of Se beam-equivalent pressure (BEP) flux, the growth temperature reduced to 400 °C with a reduced Se flux of $1\times10^{-6}$ Torr followed by 30 minutes of GaSe growth with 1:100 of Ga:Se ratio. Figure 5(b) shows the RHEED patterns along the [11$\bar{2}$0] and [10$\bar{1}$0] azimuthal orientations. Six-fold symmetry was maintained after the second low temperature step, indicating that the basal planes are aligned with the GaN substrate and there is no in-plane disorder. Figure 5(c) displays XRD spectra of grown GaSe films after the first nucleation step (black) and the second low temperature growth step (red). The GaSe layers grew along the (002) orientation, and a higher order peak (006) was observed after second step growth mainly due to the increased thickness of the film. No additional phase such as $Ga_2Se_3$ was observed after the second step growth. An off-axis $\phi$ scan of the GaSe (103) plane was performed, and six-peaks with 60° spacing were observed. The $\phi$ scan was repeated along the (102) plane of GaN and six peaks were found at the identical azimuth angles as GaSe, confirming the observation of basal plane alignment from RHEED.

Figure 6(a) shows the surface morphology of GaSe after the two-step growth process with a rms roughness of 1.1 nm. Surface coverage was found to be complete. Figure 6(b) shows the Raman spectra for GaSe grown after the first nucleation step (red), and the second low temperature step (blue). The Raman mode corresponding to a shift of 143 cm$^{-1}$ comes from the GaN/sapphire substrate. After the two-step growth, the Raman spectra matches the typical spectra expected from bulk GaSe with Raman modes at 134.3 cm$^{-1}$ ($A^1_{1g}$), 211.7 cm$^{-1}$ ($E^1_{2g}$), 250.2 cm$^{-1}$ ($E^2_{1g}$), and 307.6 cm$^{-1}$ ($A^2_{1g}$).[34] The $A^1_{1g}$ and $A^2_{1g}$ modes correspond to the out-of-plane vibration modes, while the $E^2_{1g}$ and the $E^2_{2g}$ modes are associated with the in-plane vibrational modes of GaSe. In contrast to the enhanced intensity of these Raman peaks with the film thickness, no significant peak shift of $A^2_{1g}$ mode due to the change in thickness[35] was observed because of sufficiently thick GaSe film after the first step growth. The appearance of $E^2_{1g}$ peak in GaSe has been reported in the literature.[27, 36] Nevertheless, at present the assignment of the new mode remains unclear. In addition, it is difficult to differentiate the polytypes from the Raman spectra as they show similar vibration modes.[37] Contour plot in Fig. 6(c) shows the intensity map of the dominant $A^1_{1g}$ Raman mode over a 20 μm × 20 μm area indicating complete surface coverage. Thus, this two-step growth method enables formation of coalesced multilayer GaSe films.

The microstructure of MBE-grown GaSe films were investigated in detail using STEM measurements. STEM images from two regions of the GaSe nucleation layer grown at 575 °C is shown in Fig. 7 (a) and 7 (c). An abrupt GaSe/GaN interface and 5-8 GaSe monolayers separated by van der Waals gaps could be clearly resolved in the STEM images. Ball-and-stick model generated using *VESTA* is superimposed on the atomic resolution image to identify the stacking sequence. The stacking sequence indicates that the films grown are of the $\varepsilon$-GaSe polytype, in Fig. 7(b). However, a 60° rotation of the Se-Ga-Ga-Se tetralayer is observed in the region highlighted in Fig. 7(d), in which the Ga atoms sit on top of Se atom. Figure 7(f) shows the simulated crystal structure of $\varepsilon$-GaSe with a 60° rotation of every other layer resulting in $\beta$-GaSe polytype crystal structure. Such a rotation of the basal plane would not be captured in the RHEED or XRD measurements due to the six-fold symmetry of both the $\beta$ and $\varepsilon$ polytypes of GaSe. In spite of the rotation of the first tetralayer, subsequent GaSe stacking is pure $\varepsilon$-type. This may be attributed to the fact that the $\varepsilon$ polytype is energetically more stable than the $\beta$ type.[38] Similar lattice rotations and the resultant formation of grain boundaries have been reported in the case of $MoS_2$.[39-41] Dumcenco *et al.*[41] has reported simulated data on the binding energies for $MoS_2$ and sapphire substrate as a function of orientation angle of $MoS_2$ grains. It was pointed out that only 0° or 60° orientations of the lattice were energetically favorable and stable.

The microstructure of the coalesced GaSe films grown using the two-step method was also investigated using cross-section STEM. Total number of layers after two-step growth was found to be 25-27 from STEM measurements, and 20-22 layers were grown in the second step. This implies a growth rate of 0.7 nm/min, which is similar to the low growth temperature ($T_{sub}$ = 400 $^o$C) sample. The first five layers are identical to the nucleation sample discussed in the previous section. A region with 60$^o$ rotation of first layer was also observed in the two-step sample and is shown in Fig. 8(a). However, inclusions of $\beta$-type is observed along with the dominant $\varepsilon$-type GaSe. Figure 8(b) shows a magnified image of a region cropped from the boxed region in fig. 8(a). Surface reconstruction of the GaN surface can be clearly observed in the image. Ga atoms (red arrow) at the surface are bonded directly to a Ga atom below it, suggesting a 1×1 reconstruction of Ga atoms. On top of the surface Ga atoms, two atoms (green arrows) were observed above every second Ga atom. We hypothesize that these could be Se atoms passivating the GaN surface. This suggests that van der Waals epitaxy can be used to maintain surface reconstructions on the GaN surface, and which could have important implications for Fermi level pinning and dangling bond termination at heterostructure interfaces. The electronic properties of these artificial two-dimensional interfacial layers could be of great interest, but are outside the scope of the present work. We also observed that defects formed in one area of GaSe film did not propagate along *c*-axis towards surface due to the absence of bonding between individual 2D layers (Fig. 8(c)). However, certain amount of defect propagation is indeed observed and further careful study is required to understand extended defects in 2D crystals.

CONCLUSION

In summary, we have developed a two-step method to grow continuous, crystalline films of multilayer ε-GaSe on GaN(0001). To achieve this, we first optimized the growth of GaSe films on *c*-plane sapphire and GaN(0001) substrates in the low temperature regime (optimized $T_{sub}$ = 400 °C). On both substrates, this produced continuous films of (002)-oriented GaSe with random in-plane orientation of domains. In contrast, high temperature (575 °C) growth on GaN(0001) resulted in discontinuous GaSe films, but with well-defined in-plane orientation aligned to the substrate lattice. For continuous, crystalline films, we combined these two growth modes into a two-step process where the first step is a high temperature growth to establish well-defined in-plane orientation, and the second step is a low temperature growth to coalesce the nucleated domains into a continuous film. This work illustrates the advantage of molecular beam epitaxy in realizing the growth of large area 2D crystals with high material quality.

EXPERIMENTAL METHODS

MBE growths were performed in a Veeco Gen930 system with a standard thermal effusion cell for gallium. While previous reports on GaSe growth use the standard Knudson-type effusion cell to evaporate selenium, in this work, we use a valved cracker source to supply Se. Se was evaporated using a valved cracker source with the cracker zone at 950 $^o$C in order to obtain $Se_2$ species of Se.[42-43] Growth was monitored *in-situ* using reflection high-energy electron diffraction (RHEED). Prior to the growth, *c*-plane sapphire and Fe-doped insulating GaN(0001)/sapphire substrates were solvent cleaned, annealed at 400 °C under ultra-high vacuum conditions ($1\times10^{-9}$ Torr) and loaded into the growth chamber (base pressure $7\times10^{-10}$ Torr). Sapphire substrates were then annealed at 850 $^o$C in the growth chamber for 30 minutes before ramping down the substrate temperature for GaSe growth (400-500 $^o$C). The Gallium sub-oxides on GaN substrates were removed *in-situ* prior to the growth by using the following Ga polish technique. GaN substrates were exposed to a Ga flux of $\sim1\times10^{-8}$ Torr until the RHEED showed an amorphous pattern at 400 $^o$C. The substrate was then heated to 700 $^o$C for 30 minutes, followed by a ramp down to the growth temperature. Streaky RHEED patterns with Kikuchi line patterns were obtained prior to initiation of GaSe growths on sapphire and GaN substrates. The substrate temperature was measured using the thermocouple attached with the continuous azimuthal rotation (CAR) heater. The BEP of Se was fixed at $1\times10^{-6}$ and $1\times10^{-5}$ Torr during growth and was measured using a nude ion gauge with a tungsten filament. Samples were grown at different substrate temperatures (350-600 $^o$C) and Ga:Se flux ratios. The growth was initiated by opening the Se shutter for 2 minutes followed by opening of the Ga source at the growth temperature.

The crystalline quality of the GaSe films were evaluated through X-ray diffraction (Bruker, D8 Discover) and Raman spectra (Renishaw) with a 1 mW laser at 514 nm. The surface morphology of the samples was examined by atomic force microscopy (AFM) (Veeco Instrument, DI 3000). The microstructure of GaSe was examined by cross-sectional scanning transmission electron microscopy (STEM). Due to the oxidation of GaSe in ambient conditions[44], AFM scans were performed immediately after the growth. XRD was measured after covering the GaSe surface with SPR955 photoresist. For the STEM measurements, the photoresist was removed using solvents and the surface was capped with Au metal immediately to prevent oxidation. Graphical illustrations of GaSe crystal structure was generated using *VESTA* software[45].

FIGURES

**Table of Contents Figure**

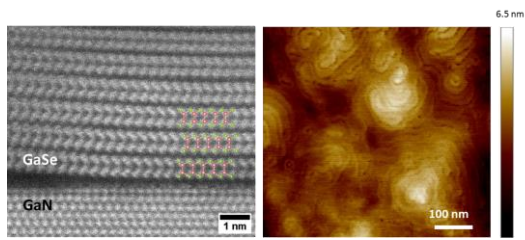

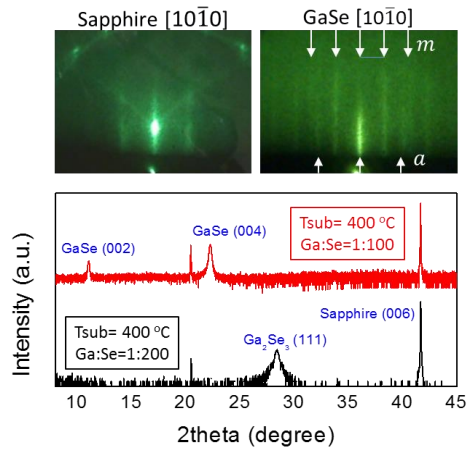

**Figure 1.** RHEED patterns observed along [10$\bar{1}$0] azimuth for (a) sapphire substrate and (b) GaSe film. (c) XRD spectra of GaSe grown at 400 $^{\circ}$C with Ga:Se of 1:200 (black) and 1:100 (red). Se flux was at maintained at $1\times10^{-6}$ Torr.

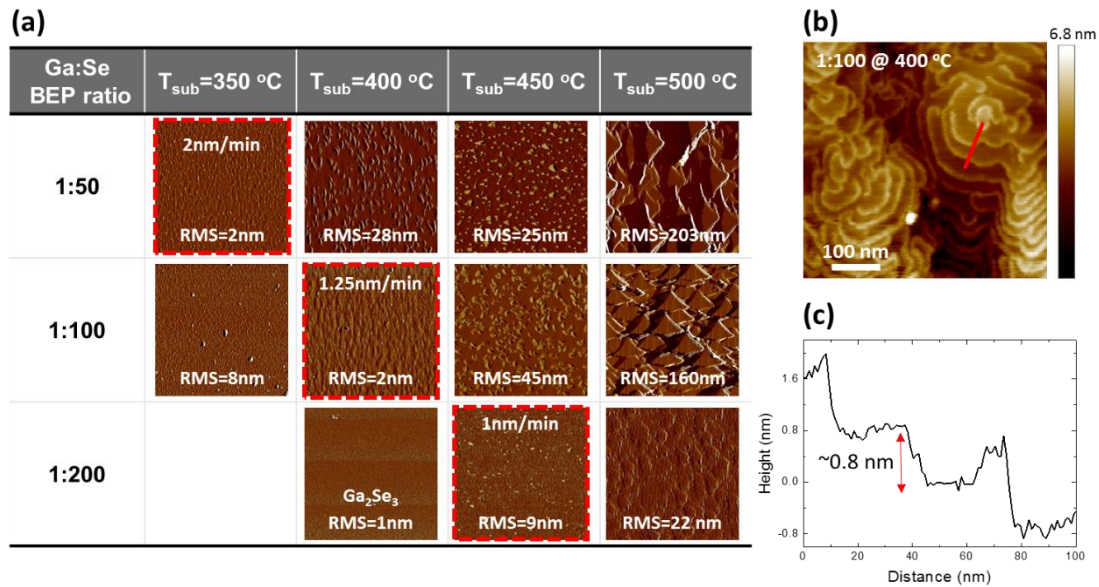

**Figure 2.** (a) AFM images of GaSe as a function of growth temperature and Ga:Se BEP flux ratio. RMS surface roughness (0.76 nm) is marked in the image. Red boxes indicate growth conditions to obtain smooth surface morphology. (b) Surface morphology of GaSe film grown at the optimized condition (400 $^{\circ}$C, Ga:Se=1:100) showing atomic steps indicating step flow growth regime. (c) Step height of GaSe film taken from the red line in (b).

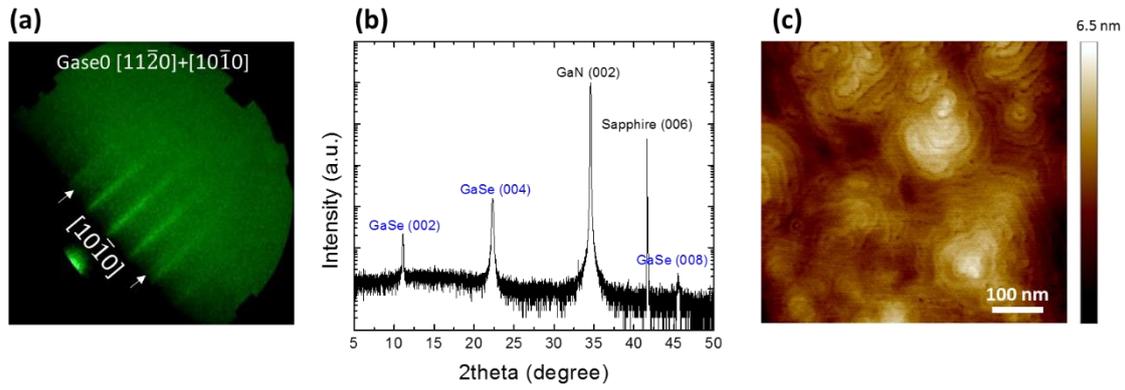

**Figure 3.** (a) RHEED pattern of GaSe showing coexistence of *a*- and *m*-planes. (b) XRD pattern, and (c) Surface morphology of GaSe film grown on GaN substrate.

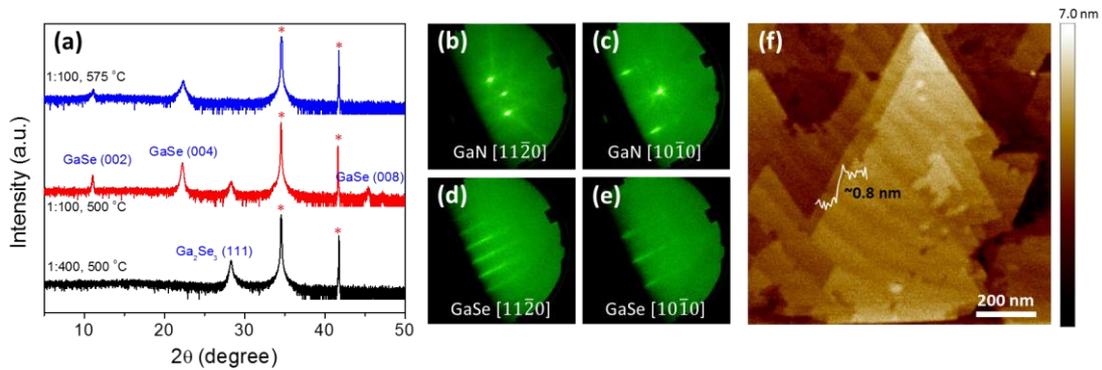

**Figure 4.** (a) XRD spectra for GaSe film grown at different condition with Se flux at $1\times10^{-5}$ Torr. The asterisks indicate the substrate peaks of GaN (002) and Sapphire (006) at 34.5 and 42 degree, respectively. (b-e) RHEED patterns of GaN and GaSe along the [11$\bar{2}$0] and [10$\bar{1}$0] azimuth showing basal plane alignment. (f) AFM image of the GaSe film showing aligned triangular domains.

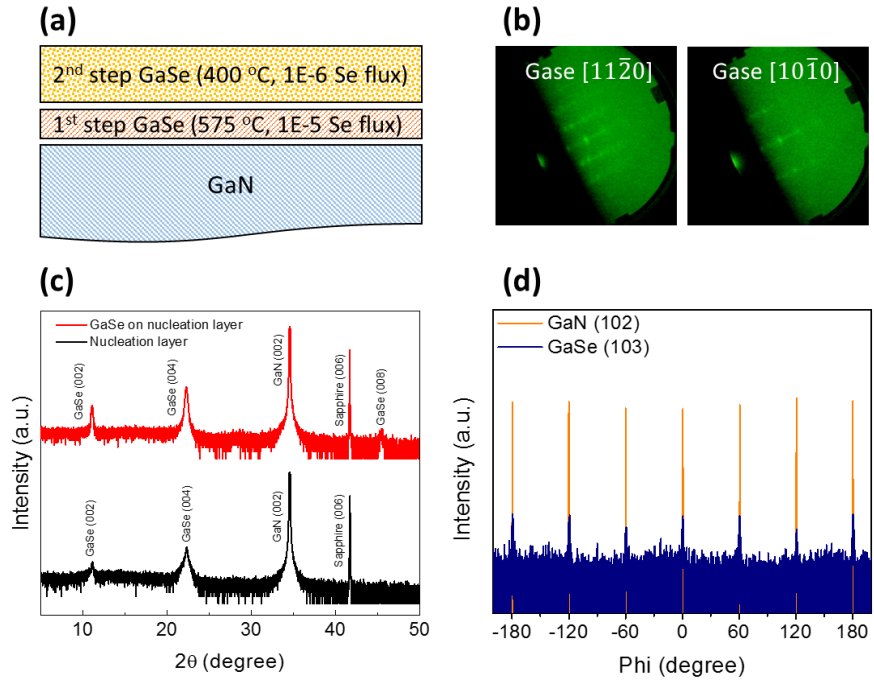

**Figure 5.** (a) Schematic of the two-step growth of GaSe on GaN substrates. (b) RHEED patterns of GaSe after the two-step growth. (c) XRD scan of GaSe after first nucleation step (black) and second (red) low temperature growth step. (d) XRD phi scan at GaSe (103) and GaN (102) planes confirming basal plane alignment.

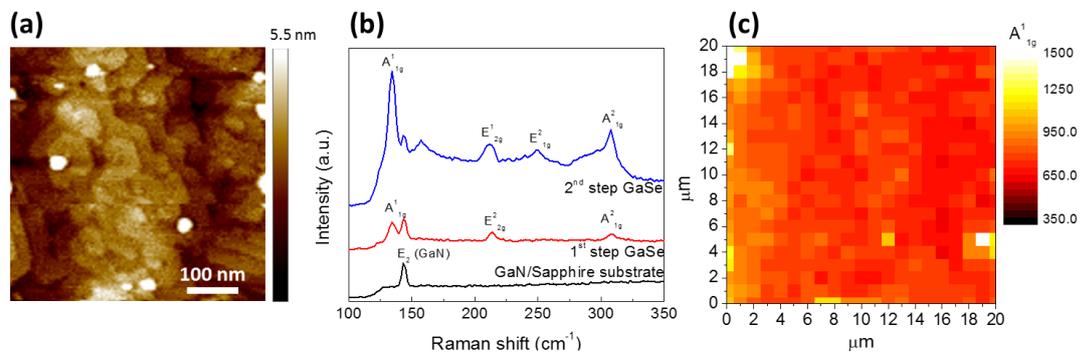

**Figure 6.** (a) AFM image of the GaSe after two-step growth. (b) Raman spectra of the GaSe film grown after first (red) and second (blue) step. Substrate is also shown for comparison. (c) Raman intensity mapping of the $A^1_{1g}$ peak over 20 μm by 20 μm.

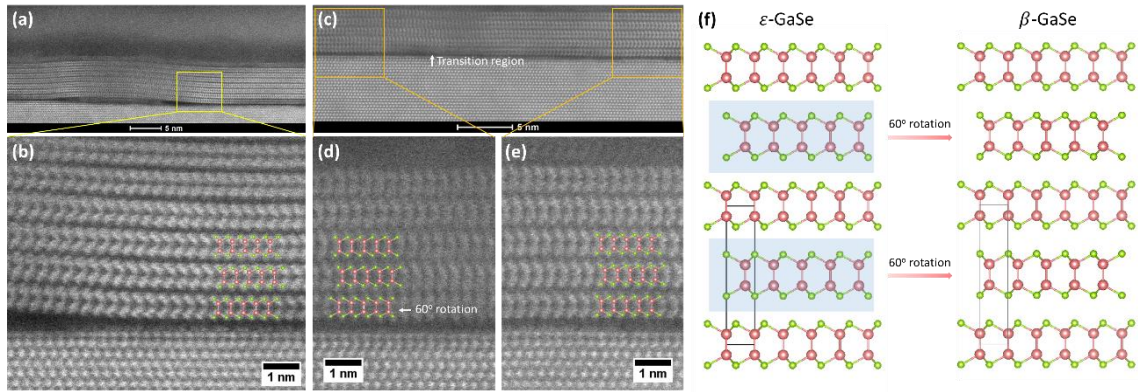

**Figure 7.** (a) Cross-sectional TEM image of GaSe film growth after first step at 575 °C. (b) Magnified image from the boxed area in (a). (c) GaSe TEM image taken from the same sample but different area. (d) and (e) Magnified images from (c). (f) Ball-and-stick model of **ε**- and **β**-GaSe types. 60° rotation of every other layer in GaSe structure in **ε**-type turns out to be **β**-type.

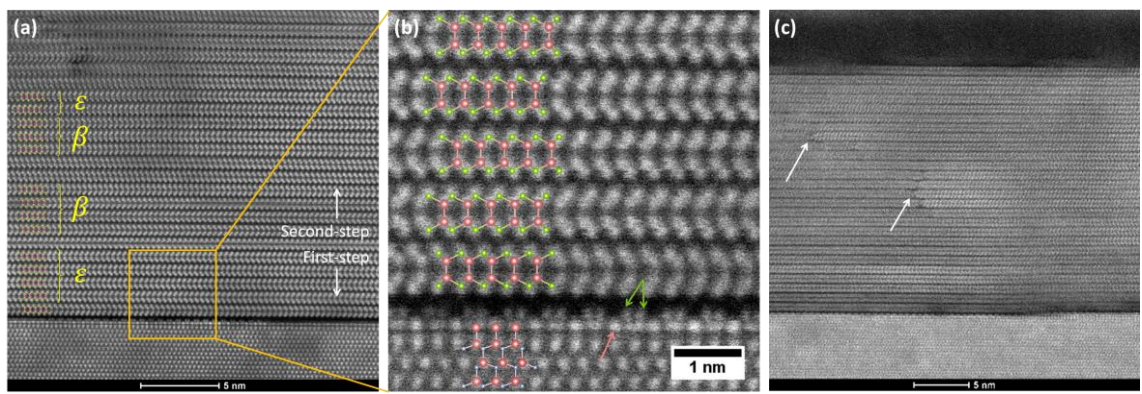

**Figure 8.** (a) and (b) Cross sectional STEM images of GaSe after two-step growth taken from different region. (c) Magnified image from boxed in (b). Ball-and-stick models of GaSe and GaN are also presented. (c) Defects formed in the middle of GaSe film marked with an arrow.

## ASSOCIATED CONTENT

**Supporting Information**. $Ga_2Se_3$ RHEED patterns, AFM scan, X-STEM, and Photo luminance spectra of GaSe. This material is available free of charge via the Internet at http://pubs.acs.org.

## AUTHOR INFORMATION

**Corresponding Author**

Email: lee.5107@osu.edu

Email: krishnamoorthy.13@osu.edu

**Author Contributions**

* Choong Hee Lee and Sriram Krishnamoorthy have contributed equally to this work

**Notes**

The authors declare no competing financial interest.

## ACKNOWLEDGMENTS

We acknowledge the support from Air Force Office of Scientific Research (AFOSR) under Contract No. FA9550-15-1-0294. The Veeco Gen930 system used for this work was purchased using NSF MRI grant (NSF DMR-1429143). RKK and DJO acknowledge the support of NSF DMR-1310661.

.

# Supplementary information of

# Molecular Beam Epitaxy of 2D-layered Gallium Selenide on GaN substrates

*Choong Hee Lee[1,*], Sriram Krishnamoorthy[1,*], Dante J. O'Hara[2], Jared M. Johnson[3], John Jamison[3], Roberto C. Myers[3], Roland K. Kawakami[2,4], Jinwoo Hwang[3], Siddharth Rajan[1]*

[1] Department of Electrical and Computer Engineering, The Ohio State University, Columbus, Ohio 43210, USA

[2] Program of Materials Science and Engineering, University of California, Riverside, CA 92521

[3] Department of Materials Science and Engineering, The Ohio State University, Columbus, Ohio 43210, USA

[4] Department of Physics, The Ohio State University, Columbus, OH 43210, USA

*Email: lee.5017@osu.edu, krishnamoorthy.13@osu.edu

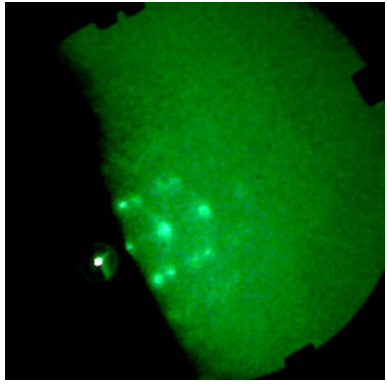

Fig. S1. Spotty RHEED patterns of $Ga_2Se_3$ grown under Se-rich condition ($T_{sub}$ = 400 $^o$C, Ga:Se = 1:200) indicating 3D growth.

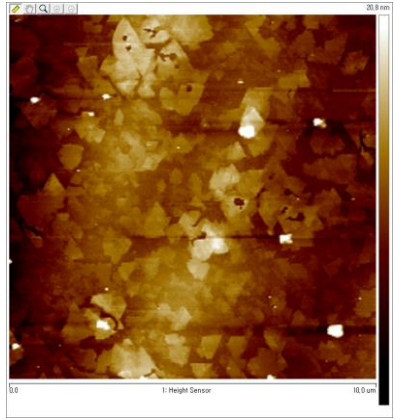

Fig. S2. Large area AFM scan (10 μm × 10 μm) image showing the discontinuity in the formation of grown GaSe layers on GaN surface at higher substrate temperature. However, the domains are found to be aligned (60$^o$ rotated domains are also visible, which is in agreement with the cross sectional STEM data).

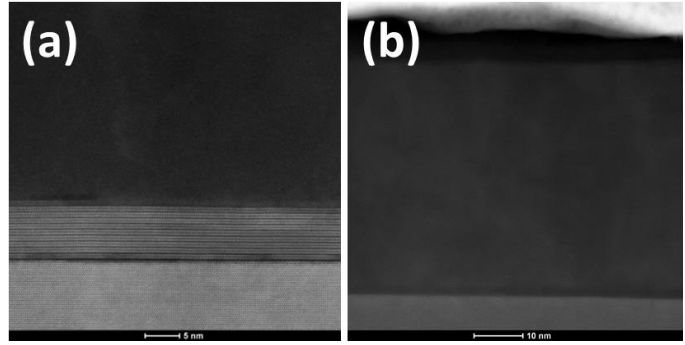

Fig. S3. Cross-sectional STEM images of GaSe nucleation layer showing discontinuous domains formed at high substrate temperature (In image S3(a) shows 9 layers of GaSe while GaSe is completely absent in image S3(b)).

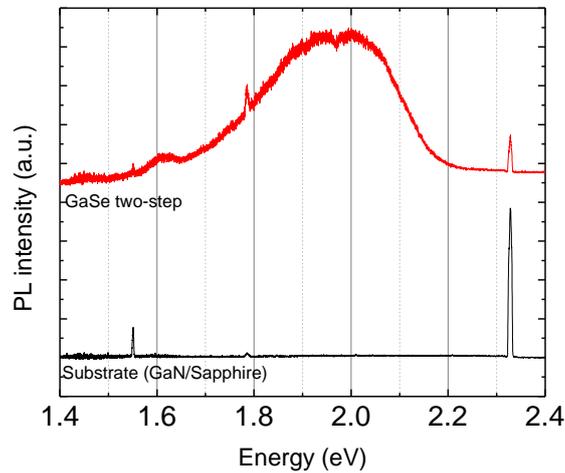

Fig. S4. Photoluminescence spectra of two-step grown GaSe film. PL spectra for Substrate (GaN/sapphire) is also included for reference.

Photoluminescence spectrum for two-step grown GaSe film is shown in Fig. S4. Pronounced emission is observed at 1.99 eV which is attributed to the band edge emission from $\varepsilon$-GaSe. We hypothesize that the peak at 1.61 eV could be arising from defects within the band gap of GaSe. The other peak observed at 1.93 eV could be originating from the mixed polytypes found in the sample.

3